\documentstyle[multicol,aps,prl,epsf,psfig]{revtex}
\begin{document}
\title{Phase Transition in a Conserved-Mass Model of Aggregation and 
Dissociation}
\author{Supriya Krishnamurthy, Satya N. Majumdar and Mustansir Barma}
\address{Tata Institute of Fundamental Research, Homi Bhabha Road, 
Mumbai-400005, India}
\maketitle

\begin{abstract}
We introduce a new model of aggregation of particles where in addition to 
diffusion and aggregation upon contact, a single unit of mass can 
dissociate from a conglomerate.  This dissociation move conserves the 
total mass and leads to a striking behaviour in the 
steady state. As the parameters are varied, the system undergoes a 
dynamical phase transition in all dimensions. In one phase the mass 
distribution decays 
exponentially for large mass whereas in the other phase there is a power 
law distribution of masses which coexists with an infinite mass aggregate.
The model is investigated analytically within mean field theory, and 
numerically in one dimension.
 
\end{abstract}

\begin{multicols}{2}
Conservation laws satisfied by the dynamics are known to modify 
drastically\cite{HH} the time-dependent behaviour of systems 
which are in thermal equilibrium.
There is, however, a wide variety of inherently {\em 
nonequilibrium} systems in nature 
whose steady states are not described by the Gibbs distribution, but are  
determined by the dynamics itself. Examples include 
systems exhibiting self-organized criticality\cite{BTW}, several
reaction-diffusion systems\cite{ZGB} and fluctuating 
interfaces\cite{edw}. Then the question naturally arises: What is the role 
played by conservation laws in
selecting the steady state of such a nonequilibrium system? The question
is particularly significant because of the propensity of many
nonequilibrium systems to organize themselves into critical states,
which are especially sensitive to such conditions. How do conservation laws
modify the power laws characteristic of such critical states? 
Can they bring in completely new types of behaviour?

We address these questions for the steady states of an important class
of nonequilibrium processes, namely those involving the twin phenomena
of aggregation and diffusion.  These processes are ubiquitous in
nature, and arise in a variety of physical settings, for instance, in
the formation of colloidal suspensions\cite{White} and polymer 
gels\cite{Ziff} on 
the one hand, and aerosols and clouds\cite{Fried} on the other.  A recent 
interesting and important result in this area due to Takayasu and 
collaborators\cite{Takayasu}, which has widespread applications, is
that constant injection of single particles into such a system leads to a 
power law distribution of particle masses in the steady state. This 
injection of particles from outside of course violates mass 
conservation. In this Letter, we 
show that a conserved-mass system in which injection is replaced by
{\em dissociation}, exhibits strikingly different and even more interesting 
behaviour: the new dissociation moves that conserve the 
total mass induce a novel dynamical phase transition. As the parameters
of the system are changed, there is a transition from one steady state 
where the mass distribution decays exponentially to another where
it decays as a power law, and  in addition develops an infinitely large
aggregate.

The conserved-mass aggregation model (CMAM) discussed here has
connections to models of gelation\cite{Ziff} and to the Takayasu model (TM)
of particle injection alluded to above\cite{Takayasu}. However it turns 
out to have a steady state structure
which is richer than either.  In experiments\cite{Paine} as well as 
theoretical models\cite{Ziff} of irreversible gel formation,
the steady-state mass distribution $P(m)$ is trivial, as there is only a
single infinite aggregate, though the kinetics of the approach to this
state is quite interesting\cite{Ziff}. In the TM on the other 
hand the injection move completely destroys the infinite aggregate and the
steady state mass distribution $P(m)$ decays as a
power law $m^{-\tau}$ for large mass, where the exponent $\tau$ depends on 
the spatial dimension.

Our conserved-mass model differs from conventional 
models of aggregation in that single
particles are allowed to chip off from more massive conglomerates.
This move corresponds to the physical process of single functional units 
breaking off
from larger clusters in the polymerization problem. It leads to a
replenishment of the lower end of the mass spectrum, and competes with
the tendency of the coalescence process to produce more massive
aggregates. The result of this competition is that two types of
steady states are possible, and there is a dynamical phase transition
between the two. In one state, $P(m)$ decays exponentially, while
the other is more interesting: $P(m)$ decays as a 
power law for large $m$ but in addition develops a delta function peak 
at $m=\infty$. Physically this means that an infinite aggregate forms 
that subsumes a finite fraction of the total mass, and coexists with 
smaller finite clusters whose mass distribution has a power law tail.
In the language of sol-gel transition, the infinite aggregate is like the
gel while the smaller clusters form the sol. However, as opposed to the 
models of irreversible gelation where the sol disappears in the steady 
state, in our model the sol coexists with the gel even in the steady state.
Interestingly, the 
mechanism of the formation of the infinite aggregate in the steady state 
resembles Bose-Einstein condensation (BEC),
though the condensate (the infinite aggregate here) forms in real space
rather than momentum space as in conventional BEC.

Our model
can be considered as the conserved counterpart of the non-conserved
TM\cite{Takayasu}. The injection move in TM that
violates the mass conservation is replaced in our model by the dissociation
move that conserves the mass. Besides being a simple model having a
self-organized critical state, TM has found widespread applications
including modeling of river networks\cite{river} and stress distribution
in granular media\cite{SNC}. It is therefore not unreasonable to expect 
that the conserved mass model discussed here will also find applications 
in a wide variety of physical processes. As an example, it can be 
considered as a simple model of river networks in a basin where there
is negligible rainfall (injection) but small rivulets can break off a
stream (dissociation).

The CMAM is defined as follows.
For simplicity we define the model on a one dimensional lattice with
periodic boundary conditions although generalizations to higher dimensions
are quite straightforward.
Beginning with a state in which the masses are placed
randomly,  a site is chosen at random. If it contains one or more than
one particle, then one of the following events can occur: 
\begin{enumerate}
\item Diffusion and Aggregation:
With probability $p_1$, the  mass $m_i$ at site $i$ moves either to
site $i-1$ or to site $i+1$.
If it moves to a site which already has some 
particles, then the total  mass just adds up.
\item Chipping (single-particle dissociation):
With probability $p_2$, a bit of the mass at the site
``chips'' off, {\it i.e.} a single particle leaves site $i$
and  moves with equal probability to one of the neighbouring
sites $i-1$ and $i+1$.
\item With a probability
$1-p_1-p_2$, the site is left undisturbed. 
\end{enumerate}
If the site chosen is empty, then nothing happens. 
The same rules 
hold even if we choose a site with only a single particle, which means
that the probability for a single particle to move left or right in
this model is $p_1 + p_2$. Note the difference with TM: in TM,
the move $2$ is replaced by addition of unit mass to every site with 
probability $1$.

We first analyze the model within the mean field approximation,
ignoring correlations in the
occupancy of adjacent sites. Then we can
directly write down
equations for $ P(m,t)$, the probability that any site
has a mass $m$ at time $t$.
\\
\begin{eqnarray}
\frac{dP(m,t)} {dt} &=& -(p_1+p_2)[1+q(t)] P(m,t) 
+ p_2 P(m+1,t)  \nonumber \\
&+&p_2 q(t) P(m-1,t)+p_1 P*P ;\;\;\; m \geq 1~~ \label{eq:mft1}\\
\frac{dP (0,t)} {dt} &=& - (p_1+p_2)q(t) P(0,t) + p_2 P(1,t) + p_1
q(t) \label{eq:mft2}. 
\end{eqnarray} 
Here $ q(t) \equiv 1-P(0,t)$ is the probability that a site is
occupied by a mass and 
$P*P=\sum_{m^{\prime}=1}^{m}P(m^{\prime},t)P(m-m^{\prime},t)$ is a
convolution term that describes the coalescence of two masses.

The above equations enumerate all possible ways in  which the  mass 
at a site might change. The first term in Eq. (\ref{eq:mft1}) is
the ``loss'' term that accounts for the probability that a
mass $m$ might move as a whole or chip off to either of the neighbouring 
sites, or a 
mass from the neighbouring site might move or chip off to the site in 
consideration. The probability of occupation of the neighbouring site, 
$q(t) = \sum_{m=1} P(m,t)$, multiplies $P(m,t)$ within the mean-field
approximation where one neglects the spatial correlations in the
occupation probabilities of neighbouring sites. The remaining three terms
in Eq. (\ref{eq:mft1}) are the ``gain'' terms enumerating the number of ways
that a site with mass $m^{\prime} \neq m$ can gain the deficit mass
$m -m^{\prime}$. The second equation Eq. (\ref{eq:mft2}) is a similar
enumeration of the possibilities for loss and gain of empty sites. 
Evidently, the mean field equations conserve the total mass.

To solve the equations, we compute the generating function, 
$Q(z,t) = \sum_{m=1}^{\infty} P(m,t)z^{m}$ from Eq. (\ref{eq:mft1}) and set 
$ \partial Q / \partial t =0$ in the steady state. We also need to use
Eq. (\ref {eq:mft2}) to write $P(1,t)$ in terms of $q(t)$. This gives
us a quadratic equation for $Q$ in the steady state.
Choosing the root that corresponds to
$Q(z=0) = 0$, we find
\\
\begin{eqnarray}
Q(z) &=&{{w+2q+wq}\over {2}}-{w\over {2z}}-{wqz\over {2}} 
\nonumber \\
&+& wq{(1-z)\over {2z}}\sqrt {(z-z_1)(z-z_2)}.  \label{eq:qsol}
\end{eqnarray}
where $w=p_2/p_1$ and $z_{1,2}=(w+2\mp 2\sqrt {w+1})/wq$.
The value of the occupation probability $q$ is
fixed by mass conservation which implies that $\sum mP(m)=M/L\equiv \rho$.
Putting ${\partial}_zQ (z=1)=\rho$, the resulting relation between $\rho$ 
and $q$ is \begin{equation}
2\rho = w(1-q) - wq\sqrt {(z_1-1)(z_2-1)}~.
\label{eq:defq}
\end{equation}

The steady state probability distribution $P(m)$ is the coefficient of 
$z^m$ in $Q(z)$ and can be obtained from $Q(z)$ in Eq. (\ref{eq:qsol})
by evaluating the integral 
\begin{equation}
P(m) = {1\over {2\pi i}}\int_{C_o} \frac {Q(z)} {z^{ m+1}} dz
\label{eq:contour}
\end{equation}
over the contour $C_o$ encircling the origin. 
The singularities of the integrand govern the asymptotic behaviour of 
$P(m)$ for large $m$. Clearly the integrand has branch cuts at 
$z=z_{1,2}$. For fixed $w$, if one increases the density $\rho$, the
occupation probability $q$ also increases as evident from Eq. 
(\ref{eq:defq}).
As a result, both the roots $z_{1,2}$ start decreasing. As long as the
lower root $z_1$ is greater than $1$, Eq. (\ref{eq:defq}) is well defined
and the analysis of the contour integration around the branch cut
$z=z_1$, yields for large $m$,
\begin{equation}
P(m) \sim e^{-m/{m^{*}}}/m^{3/2} ~,
\end{equation}
where the characteristic mass, $m^{*}=1/{\log (z_1)}$ and diverges as
$\sim (q_c-q)^{-1}$ as $q$ approaches $q_c =(w+2-2\sqrt {w+1})/w$. 
$q_c$ is the critical value of $q$ at which $z_1=1$. This exponentially 
decaying mass distribution is the signature of ``disordered" phase which 
occurs for $q<q_c$ or equivalently from Eq. (\ref{eq:defq}) for
$\rho < {\rho}_c={\sqrt {w+1}} -1$.

When $\rho={\rho}_c$, we have $z_1=1$, and analysis of the contour around 
$z=z_1=1$ yields a power law decay of $P(m)$,
\begin{equation}
P(m)\sim m^{-5/2}.
\end{equation}
As $\rho$ is increased further beyond ${\rho}_c$, $q$ can not increase
any more because if it does so, the root $z_1$ would be less than $1$ 
(while the other root $z_2$ is still bigger than $1$) and
Eq. (\ref{eq:defq}) would be undefined. The only possibility is that $q$
sticks to its critical value $q_c$ or equivalently the lower root $z_1$
sticks to $1$. Physically this implies that adding more particles   
does not change the occupation probability of sites. This can happen only 
if all the additional particles (as $\rho$ is increased) aggregate on a 
vanishing fraction of sites, thus
not contributing to the occupation of the others. Hence in this 
``infinite-aggregate" 
phase $P(m)$ has an infinite-mass aggregate, in addition to the power law
decay $m^{-5/2}$. Concomitantly Eq. (\ref{eq:defq}) ceases to hold, and 
the relation now becomes 
\begin{equation}
\rho = {w\over {2}}(1-q_c) + \rho_{\infty}
\end{equation}
where $\rho_{\infty}$ is the fraction of the mass in the infinite aggregate.
The mechanism of the formation of the aggregate is reminiscent of Bose 
Einstein condensation. In that case, for temperatures in which a macroscopic
condensate exists, particles added to the system do not contribute to the 
occupation of the excited states; they only add to the condensate, as 
they do to the infinite aggregate here.

Thus the mean field phase diagram (Fig. 1) of the system consists of 
two phases,
``disordered" and ``infinite-aggregate", which are separated by the
phase boundary, $\rho_c={\sqrt {w+1}}-1$. While this
phase diagram remains qualitatively the same even in $1$-d, 
the exponents characterizing the power laws are different from
their mean field values (see Fig. 1).


We have studied this model using Monte Carlo simulations on a
one-dimensional lattice. Although we present results here for a 
relatively small 
size lattice, $L=1024$, we have checked our results for larger sizes
as well. We 
confirmed that all
the qualitative predictions of the mean-field theory remain true, by
calculating $P(m)$ numerically in the steady state.  Figure 2 
displays two numerically obtained plots of
$P(m)$. The existence of both the disordered (denoted by $+$)
and the infinite-aggregate phase (denoted by $\times $) is confirmed. 
In particular, the second curve shows clear evidence of a power-law behaviour
of the distribution, which is cut off by finite-size effects, and
for an `infinite' aggregate beyond. We confirmed that the mass $M_{agg}$
in this aggregate grows linearly with the size, and that the
spread $\delta M_{agg}$ grows sublinearly, implying that the ratio
$\delta M_{agg}/M_{agg}$ approaches zero in the thermodynamic limit.
The exponent $ \tau $ which characterizes the finite-mass fragment power 
law decay is numerically found to be $ 2.33\pm .02$. The difference of this 
value from the mean-field value 2.5 is presumably due to the
neglect of spatial correlations within mean-field theory.



We note that in TM, the exponent $\tau_{TM}$ that characterizes the power
law decay of mass distribution in the steady state has an exact value
$4/3$ in $1$-d and $3/2$ within mean field theory\cite{Takayasu}. In the 
CMAM, we 
find that in the aggregate phase $\tau_{CMAM}\sim 2.33$ in $1$-d and $5/2$
within the mean field theory. It is therefore tempting to conjecture that 
$\tau_{CMAM}=\tau_{TM}+1$ although we have no proof of this.

Mass conservation evidently affects the steady state in this class of
nonequilibrium models, but
what are the other factors which determine the universality classes?
We addressed this question by studying the effect of directionality
on the motion of the masses, leading to a finite mass current. For the 
mass-nonconserved TM, it is known that making such
a change has no effect on the scaling properties\cite{Takayasu}. Our 
numerical study of the directed mass-conserved case shows\cite{KMB} that 
directionality in fact changes the universality class. As in the 
undirected case, the model continues to show a phase transition between 
two phases, without and with an infinite aggregate, respectively. However,
the exponent $\tau$ characterizing the power law decay in the 
infinite-aggregate phase, is different in this case (Fig. 3).
In this model, $\tau \simeq 2.05$ within numerical error\cite{KMB}.
Clearly, $P(m)$ 
should decay faster than $m^{-2}$ for large $m$ to keep the total mass
$ \sum m P(m)$ finite. Perhaps
$P(m)$ decays as $m^{-2}$ with additional logarithmic factors.



Interestingly, the CMAM can be mapped onto a driven diffusive lattice gas
model, and thence to a model of interface dynamics, so that our results for
aggregation phenomena have wider significance.
A configuration of the CMAM is mapped onto a particle-hole configuration as 
follows: Every site with $m$ particles in the aggregation model
is mapped to a cluster of $m$ successively
occupied sites, with a vacancy at the rightmost edge in the particle
model. An empty site maps to an empty site. Thus a lattice of $L$ sites in
the aggregation-dissociation model maps to one of $L+N_p$ sites in the
hard-core particle model, where $N_p$ is the total number of particles.
The dynamics defined in the earlier section now translates to the
following rules for particle motion.
If a randomly selected site is occupied, then with probability $p_1$,
the entire particle cluster to which the occupied site belongs is moved
one site to the left or right. With probability $p_2$, the
rightmost or leftmost particle from the cluster dissociates from the
cluster and moves one site to the right or left. With probability
$1-p_1-p_2$, the system is left undisturbed.
The mapping to an interface model follows a standard route\cite{sep}:
a particle is mapped to a unit segment with a positive slope, while a 
hole maps to a segment of unit length with a negative slope. The 
equivalent interface then allows  both for single corner flips at 
hills and valleys, and also for slice-wise moves of connected segments of
up and down slopes.
Under this dynamics, interface profiles can
develop  very different spatial structures as compared to
those described in customary growth  processes \cite{edw}.
In the exponential phase, particle cluster lengths decay 
exponentially as is usual, but in the infinite-aggregate phase,
a single macroscopically long stretch with positive slope develops. This 
unusual behaviour differs from descriptions of mound formation\cite{Krug},
in that the infinite stretch is 
accompanied by the formation of fairly long finite stretches with
a power-law distribution of stretch lengths. This defines a new 
universality class in interface dynamics.

In conclusion, we point out that several
questions still remain open. Amongst these is the understanding of
spatial and temporal correlations, and
the role of directionality and dimensionality in influencing the scaling 
behaviour, in particular, the determination of the upper critical
dimension.

We thank Deepak Dhar for useful discussions.

\end{multicols}

{\bf {Figure Captions}}

Fig. 1. The phase diagram in the $\rho$-$w$ plane.
The dashed line shows the mean field phase boundary. The points 
denoted by ($\bullet$) are obtained numerically in one dimension. 

Fig. 2. The mass distribution $P(m)$ vs. $m$ for the undirected model 
in $1$-d on a log-log plot for $w=1.0$ and $\rho=0.2$ (shown by $\times$ 
symbols) and for $w=1.0$ and $\rho=3.0$ (shown by $+$ symbols). 

Fig. 3. The mass distribution $P(m)$ vs. $m$ (only the power law part) on a 
log-log plot for both undirected ($+$) and directed ($\times$) case in 
the ``Infinite-Aggregate" phase at density $\rho=12.0$ and $w=1.0$. The 
straight line with slope $-2$ is drawn as a guidance to the eyes.


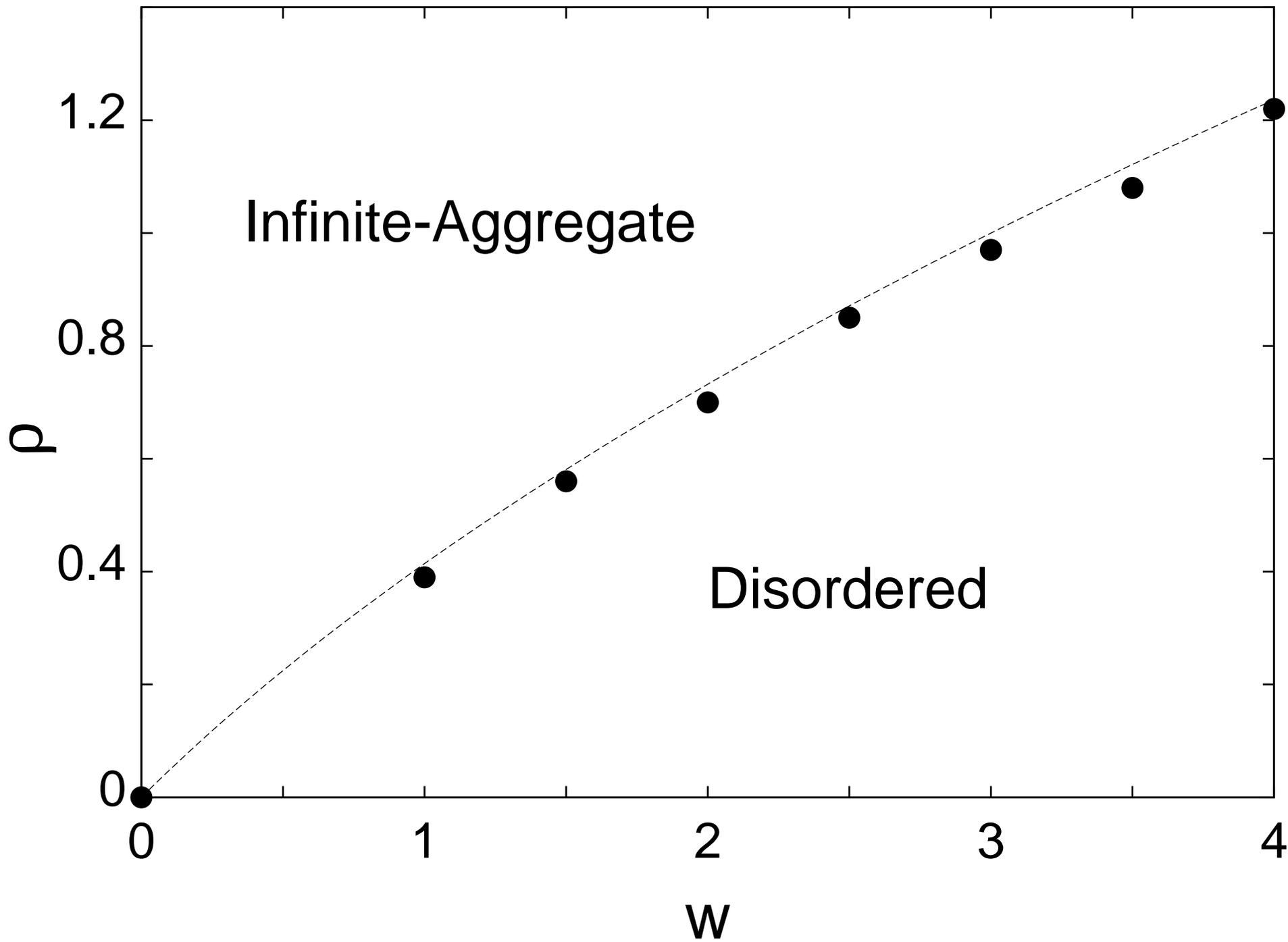
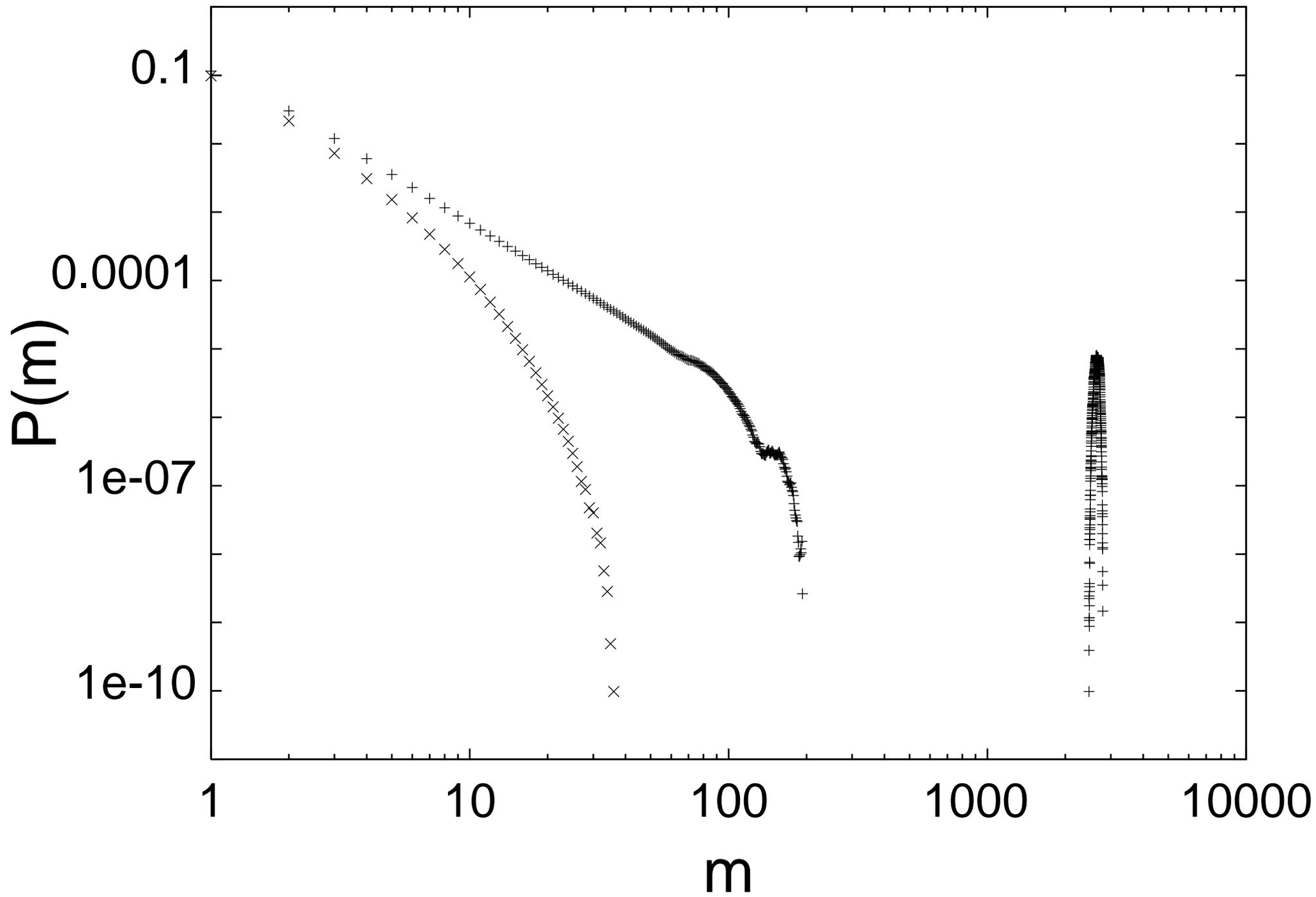
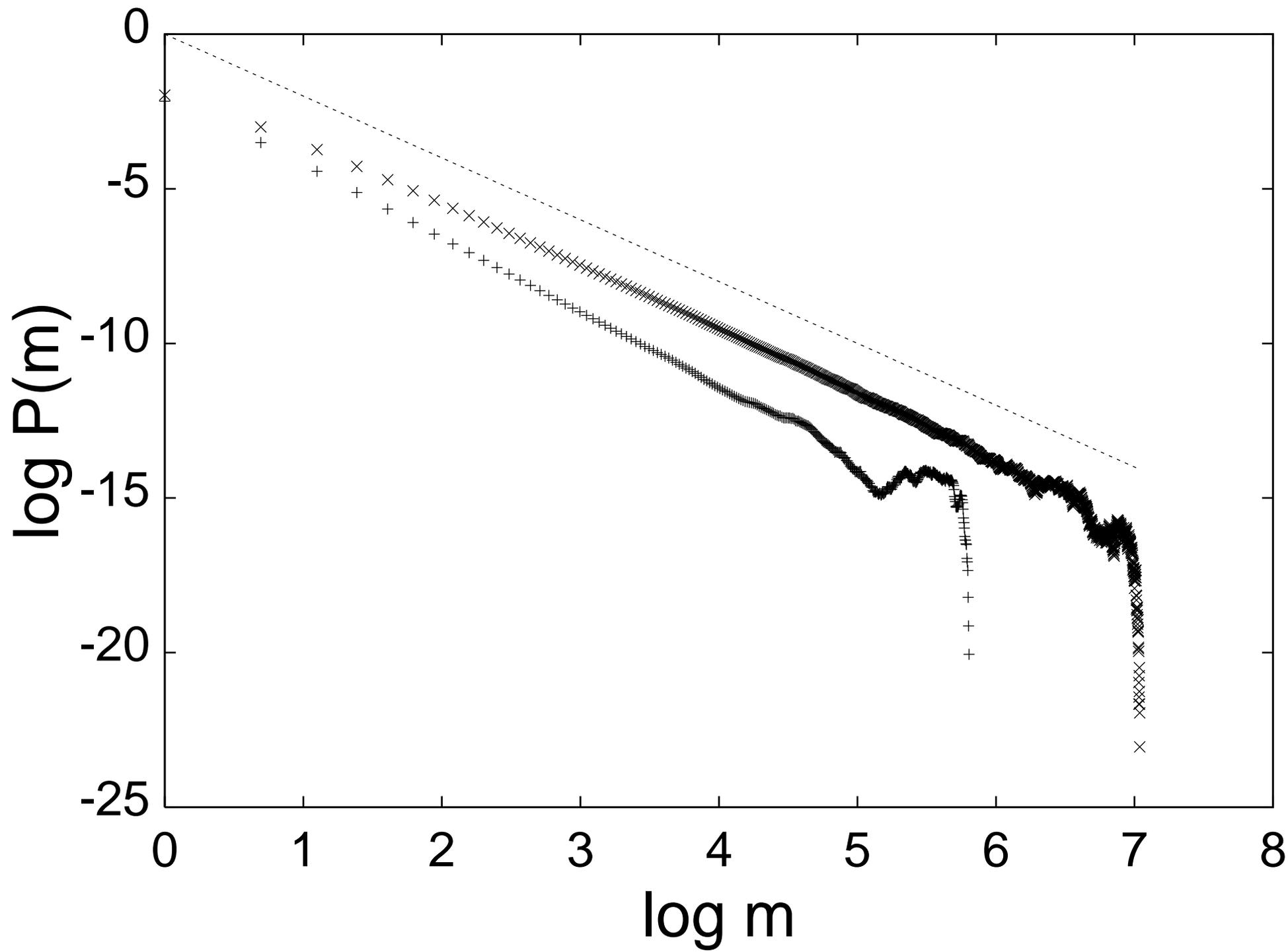
\end{document}